\documentclass[epj,times,nopacs]{svjour}
\usepackage{ifpdf}
\usepackage{multirow}
\usepackage[numbers]{natbib}
\usepackage{xspace}
\usepackage{graphicx}
\usepackage{amsmath}
\usepackage{amssymb}
\usepackage{color}
\usepackage[normalem]{ulem}	

\newcommand{\de}{$\Delta$E\xspace}
\newcommand{\dee}{$\Delta$E-E\xspace}

\newcommand{\eres}{E$_{\mathrm{RES}}$\xspace}

\newcommand{\fom}{FoM\xspace}

\newcommand{\adc}{ADC\xspace}

\newcommand{\fazia}{FAZIA\xspace}

\newcommand{\psa}{PSA\xspace}

\title{Energy measurement and fragment identification using digital signals from partially depleted Si detectors}

\author{G.Pasquali\inst{1,2\mail{pasquali@fi.infn.it}}
\and G.Pastore\inst{1,2}
\and N.Le Neindre\inst{3}
\and G.Ademard\inst{4}
\and S.Barlini\inst{1,2}
\and M.Bini\inst{1,2}
\and E.Bonnet\inst{5}
\and B.Borderie\inst{4}
\and R.Bougault\inst{3}
\and M.Bruno\inst{6}
\and G.Casini\inst{2}
\and A.Chbihi\inst{5}
\and M.Cinausero\inst{7}
\and J.A.Due{\~n}as\inst{8}
\and P.Edelbruck\inst{4}
\and J.D.Frankland\inst{5}
\and F.Gramegna\inst{7}
\and D.Gruyer\inst{5}
\and A.Kordyasz\inst{9}
\and T.Kozik\inst{10}
\and O.Lopez\inst{3}
\and T.Marchi\inst{7}
\and L.Morelli\inst{6}
\and A.Olmi\inst{2}
\and A.Ordine\inst{11}
\and M.P\^arlog\inst{3,12}
\and S.Piantelli\inst{2}
\and G.Poggi\inst{1,2}
\and M.F.Rivet\inst{4}
\and E.Rosato\inst{11,13}
\and F.Salomon\inst{4}
\and G.Spadaccini\inst{11,13}
\and A.A.Stefanini\inst{1,2}
\and S.Valdr\`e\inst{1,2}
\and E.Vient\inst{3}
\and T.Twar\'og\inst{10}
\and R.Alba\inst{14}
\and C.Maiolino\inst{14}
\and D.Santonocito\inst{14}\\\\
\textbf{For the FAZIA Collaboration}
}

\institute{Dipartimento di Fisica, Universit\`a di Firenze, via G.Sansone 1, 50019 Sesto Fiorentino (FI), Italy
\and INFN Sezione di Firenze, via G.Sansone 1, 50019 Sesto Fiorentino (FI), Italy
\and LPC, IN2P3-CNRS, ENSICAEN et Universit\'e de Caen, F-14050 Caen-Cedex, France
\and Institut de Physique Nucl\'eaire, CNRS/IN2P3, Universit\'e Paris-Sud 11, F-91406 Orsay cedex, France
\and GANIL, CEA/DSM-CNRS/IN2P3, B.P. 5027, F-14076 Caen cedex, France
\and INFN and Universit\`a di Bologna, 40126 Bologna, Italy
\and INFN-LNL Legnaro, viale dell’Universit\`a 2, 35020 Legnaro (Padova) Italy
\and Departamento de Fisica Aplicada, FCCEE Universidad de Huelva, 21071 Huelva, Spain
\and Heavy Ion Laboratory, University of Warsaw, ul. Pasteura 5a, 02-093 Warsaw, Poland
\and Jagiellonian University, Institute of Nuclear Physics IFJ-PAN, PL-31342, Krak´ow, Poland
\and INFN Sezione di Napoli, 80126 Napoli, Italy
\and ``Horia Hulubei'' National Institute of Physics and Nuclear Engineering, RO-077125 Bucharest, Romania
\and Dipartimento di Fisica, Universit\`a di Napoli ”Federico II”, 80126 Napoli, Italy
\and INFN-LNS Catania, 95129 Catania, Italy
}

\abstract{
A study of identification properties of a Si-Si \dee telescope exploiting an underdepleted residual-energy detector
has been performed. Five different bias voltages have been used, one corresponding to full depletion, the others
associated with a depleted layer ranging from 90\% to 60\% of the detector thickness.
Fragment identification has been performed using either the \dee technique or Pulse Shape Analysis (PSA).
Both detectors are reverse mounted: particles enter from the low field side, to enhance the PSA performance.
The achieved charge and mass resolution has been quantitatively expressed using a Figure of Merit (\fom).
Charge collection efficiency has been evaluated and the possibility of energy calibration corrections has
been considered.
We find that the \dee performance is not affected 
by incomplete depletion even when only
60\% of the wafer is depleted. Isotopic separation capability
improves at lower bias voltages with respect to full depletion, though charge identification thresholds are higher 
than at full depletion. Good isotopic identification via PSA has been obtained from a
partially depleted detector whose doping uniformity is
not good enough for isotopic identification at full depletion.
}

\begin{document}

\authorrunning{G.Pasquali \textit{et al.}}
\titlerunning{Energy measurement and fragment identification  from partially depleted Si detectors}
\maketitle

\section{Introduction}

In recent years, intensive experimental work has been devoted
to improving the nuclear fragment identification techniques based on Pulse Shape (hereafter PS; PSA for PS Analysis) 
 applied to Si detector signals.
In fact, identification in mass (A) and charge (Z) of light charged particles and intermediate mass fragments  
will be particularly useful
at Radioactive Ion Beam facilities for studies focused on nuclear isospin,
where the N/Z ratio of the products will be a key experimental observable~\cite{n_su_z1, n_su_z2, n_su_z3, n_su_z4}.

Large solid angle detector arrays usually feature a \dee telescope as elemental cell~\cite{indra, chimera, garfield, nimrod}. A \dee telescope is a
multi-layer detection system: the  impinging particle passes through the detectors one after the other and the energy
deposited in each detector is measured~\cite{bromley}.
 However, neither fragments stopped in the first \de detector nor fragments punching through
 the whole telescope can be uniquely identified. 
On the other hand, in a Si-Si-CsI telescope like those developed by the \fazia collaboration~\cite{fazia},
PSA would allow identification of fragments stopped in the first Si detector, thus  considerably lowering
the energy threshold for identification.

\begin{table*}[htbp]
 \begin{center}
\begin{tabular}{l c c c}
\hline 
\\
 Technical Specs. & Si1 & Si2 & CsI(Tl)   \\ 
\\
\hline \\

Manufacturer & FBK  &   FBK   &  Amcrys  \\
Bulk Type & n  &   n   &    \\
Thickness & 311$\,\mu$m & 510$\,\mu$m & 10 cm+FBK diode\\
Active Area & 20$\times$20 mm$^2$ &  20$\times$20 mm$^2$ & 21$\times$21 mm$^2$ \\
Depletion Voltage & 140~V & 290~V & \\
Applied Voltage & 140~V & 105-290~V & 30~V \\
Resistivity ($\Omega\,$cm) & $\sim\, 2550$  & $\sim\, 2900$ & \\
Resistivity Uniformity (FWHM) & $\sim\, 4$\%  & $\sim\, 6$\% & \\
Carrier Lifetime ($\mu$s)     &       6000   &  6000     & \\
Digitizer (bit/rate) & 14/100MHz& 14/100MHz& 12/125MHz \\ 
Digitizer board ENOB  &  11.4& 11.4& 10 \\ 
Energy Full Scale (Si-GeV) &  3.7  & 2.5  &  0.4  \\
P.A. Decay Const. ($\mu$s)   & 750 & 425 &\\
Acquired signal length  ($\mu$s)  & 20   & 70   &  30 \\
Trapezoidal Shaper Rise Time ($\mu$s)   &  2  &  2  &   \\
Trapezoidal Shaper Flat Top ($\mu$s)   &  1  &  55  &   \\
Gain (keV/LSB)  &  283   & 192  &  \\
\\
\hline
\end{tabular}
\end{center}
\caption{Main features of the telescope employed in our test.  The depletion voltage has been obtained from
the C-V characteristics of the detectors. ``Applied voltage'' means
the actual value applied to the silicon detector, taking into account the voltage drop on the bias resistor due to the leakage current.
The full scale energy value
takes into account the position of the signal baseline in the \adc range. Gain is given in keV/LSB units, where LSB is
the Least Significant Bit of the \adc.} 
 \label{tab:detectors}
\end{table*}

\begin{figure}[htbp]
  \centering
\includegraphics[width=8cm]{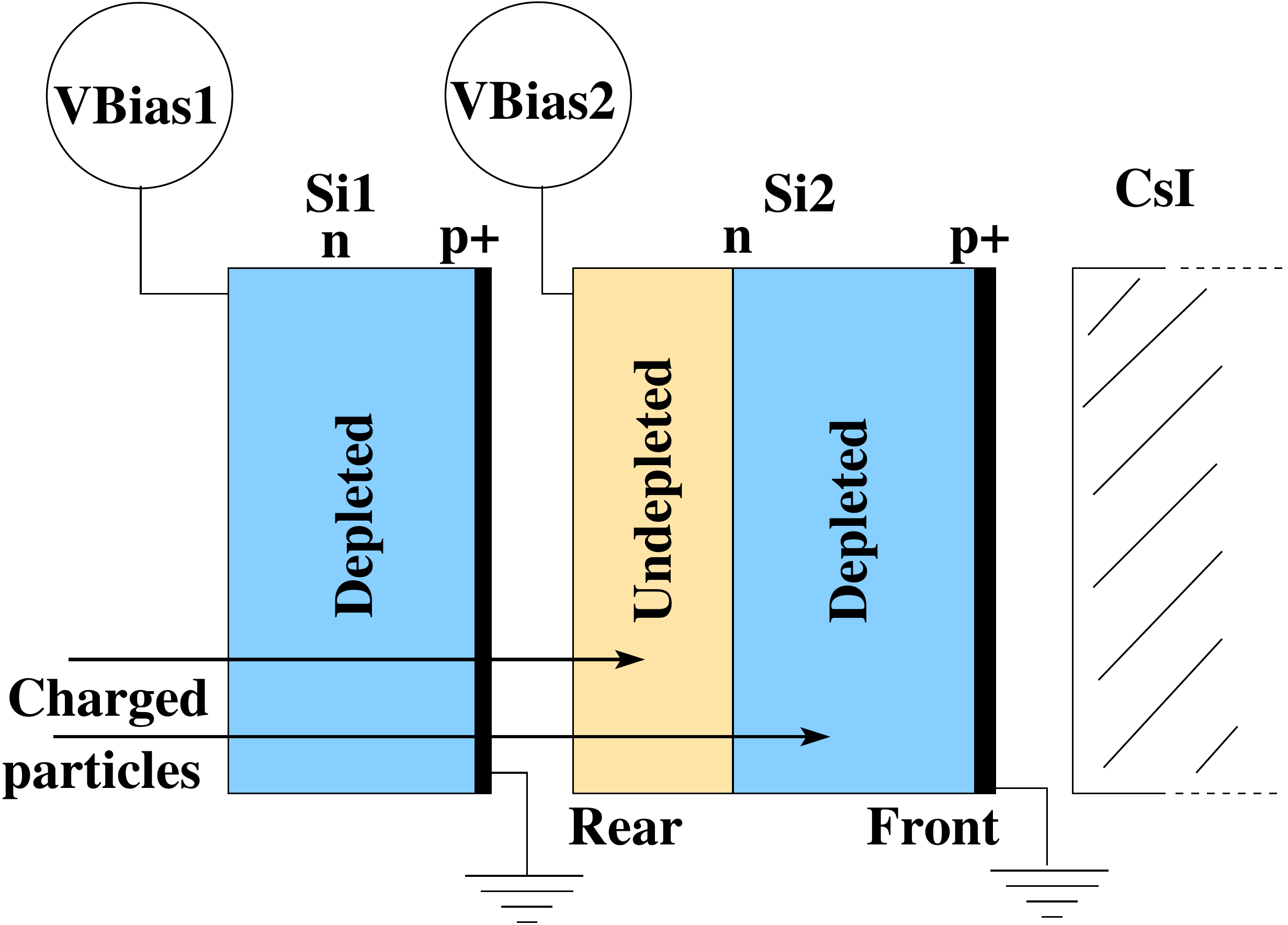}
   \caption{(color online) Sketch of the \dee telescope employed in this work. 
   Si detectors are mounted with the low field side facing the target (rear side injection).
   When Si2 is not fully depleted, particles enter the detector from the undepleted region.
   The picture is not to scale.}
   \label{fig:telescope}
\end{figure}

Much has been learnt from previous tests of \fazia telescope 
prototypes~\cite{bardelli_lnl07, carboni_lns09, leneindre_lns11, sct_gab2012}.
Other tests have been performed in the framework of the NUCL-EX collaboration, exploiting
the GARFIELD+RCo detector array at LNL~\cite{bruno_garf}. 
An improvement of PSA isotopic identification capabilities in partially depleted Si detectors has been
observed. A more systematic study of the PS identification capabilities of underdepleted 
Si detectors has thus been started. In a  test performed at Laboratori Nazionali del Sud (LNS) of  INFN,
Si2, the second Si stage of a standard \fazia telescope (see Fig.~\ref{fig:telescope}), has been biased at five different bias
voltages. One of the employed voltages corresponds to full depletion, the others are
associated with depletion thicknesses ranging from 90\% to 60\% of the detector thickness.
Though in a physics experiment one would obviously  employ
PSA on the first stage to lower the identification thresholds~\cite{carboni_lns09, leneindre_lns11},
studying the second stage is a school case which has many advantages. For instance,
fragments stopped in the second stage can  be identified with the
\dee technique exploiting the correlation between the charge collected from the
two Si detectors. Knowing the fragment charge and mass, it is possible, \textit{e.g.},
to estimate its incident energy from the \de energy deposited in the first stage, Si1. 
The energy value obtained from the calibration of the second stage can thus be cross-checked with the
estimate based on the first one. PSA performance can also be better studied
since charge and mass of the fragments stopped in Si2 are known from the \dee
correlation. The last statement also applies  to partially depleted detectors since
the quality of \dee identification is unchanged (see sect.~\ref{subsec:dee}).

To improve PS identification capabilites, the
 Si detectors of \fazia telescopes are mounted with the ohmic (low field) side facing the target~\cite{leneindre_lns11}.
Therefore, fragments enter a partially depleted detector from the undepleted region.
In this work, we were interested in the following issues: 

\begin{itemize}
 \item what is the charge collection efficiency in the
partially depleted detector for particles impinging on the undepleted region?
\item  Is
 the detector energy response linear with the deposited energy in such conditions? 
\item Is the \dee identification affected by incomplete depletion
of the detector?
\item Is there really an improvement in PS identification?
\item What are the energy thresholds for charge and mass identification? 
\end{itemize}

A doping uniformity of about 1\% FWHM, or less, would be needed to discriminate, \textit{e.g.}, 
carbon isotopes via PSA in totally depleted detectors~\cite{bardelli_lnl07}.   
The doping uniformity of the detector under test is about 6\%.  
It does not allow carbon isotopic identification via PSA 
at full depletion voltage (see sect.~\ref{subsec:psa}) but it does when it is
underdepleted, even though in a reduced energy domain. 
A recent paper~\cite{duenas13} showed that proton-deuteron separation at low energies improves when working near depletion
voltage with respect to overdepletion. However,
 our work reports for the first time as far as we know,  about the possibility of improving
the identification via PSA of fragments covering a relatively large range of charge and mass (and impinging energy)
by underdepleting the silicon detector.
 

In section~\ref{sec:setup} the employed experimental setup  is illustrated together with detector signal treatment.
Section~\ref{sec:anal} discusses the performance achie\-ved at the various
bias voltages. In particular, section~\ref{subsec:ampli} presents the problem of charge amplitude
estimation with the extremely slow signals coming from an underbiased detector,
section~\ref{subsec:dee}  deals with the particle identification capabilites
of the underdepleted detector using the usual \dee technique in a telescope configuration.
Section~\ref{subsec:energy} deals with energy calibration using the so-called ``punch-through'' points.
In section~\ref{subsec:calib} the dependence of charge collection efficiency on the particle range
is discussed. 
Finally, section~\ref{subsec:psa} shows how fragment identification using PS identification methods
depends on the applied bias voltage. Both the quality of charge and mass separation  and the 
energy thresholds for identification are considered.

\section{Experimental setup} \label{sec:setup}

The data presented in this work were collected in Catania  at
Laboratori Nazionali del Sud (LNS) of  INFN. The beam was
 $^{84}$Kr at 35A MeV impinging on  $^{112}$Sn  and $^{197}$Au targets
of thickness 488~$\mu$g/cm$^2$ and 178~$\mu$g/cm$^2$, respectively.

A telescope composed of Si 300~$\mu$m - Si 500~$\mu$m - CsI(Tl) 10~cm, was mounted
in the ``Ciclope'' scattering chamber, at a distance from
target of 50~cm and at $\approx\,8^\mathrm{o}$ polar angle with respect to the
beam line, covering about $2^\mathrm{o}$ in polar angle. A sketch of the telescope is
shown in Fig.~\ref{fig:telescope} where two
kinds of events are also deicted (arrows): particles stopped in the undepleted region of Si2 and
particles reaching the depleted region.

The angular position of the telescope was
 slightly beyond the grazing angle, where the reaction mechanism concentrates most of the 
products (the grazing angle was $\sim4.1^\mathrm{o}$ and $\sim6.0^\mathrm{o}$
 for $^{112}$Sn and $^{197}$Au targets, respectively). The same kind of telescope was
employed in~\cite{leneindre_lns11}. In Table~\ref{tab:detectors} we summarize
the main characteristics of the telescope and of the
dedicated front-end electronics (FEE). The detector thickness was measured via
a precision gauge with an uncertainty of $\pm\,1\mu$m and it was found to be 311~$\mu$m
for Si1 and 510~$\mu$m for Si2.

According to the usual FAZIA recipe for optimizing the PSA performance,
both silicon detectors, manufactured by FBK (Trento, Italy)~\cite{fbk}, were of the
neutron transmutation doped (n-TD) type (for a better doping uniformity~\cite{ bardelli_lnl07, bardelli_laser})
 and cut at a ``random'' direction (to avoid ``channeling'' effects~\cite{bardelli_lnl07, bardelli_channeling}). 
The mechanical mounting allowed us to use the Si detectors in transmission.

%

The voltage applied to the Si detectors was kept constant at the desired value using
a bias system with reverse current monitoring that takes into account the voltage
drop on the bias resistors of the preamplifiers (20~M$\Omega$). However, no
substantial correction was needed during data taking, since no increase in
the reverse current (always less than 50~nA) has been noticed.

The bias voltage of Si1 has been kept at 140~V, full depletion value, during the whole
measurement. Five different bias voltages have been applied to Si2, acquiring more
than $5\times 10^5$  events  in each case.  The number of events collected for each bias voltage is
reported in Table~\ref{tab:Si2bias}, rightmost column.  Table~\ref{tab:Si2bias} also reports
the  estimated depletion depth and maximum measured rise-time at all  employed bias voltages
(all  rise-times quoted in this paper are taken from 20 to 70\% of the maximum charge signal amplitude).
The estimate of the maximum rise-time reached at each bias voltage has been obtained from
the ``Energy vs Charge rise-time'' correlations (see, \textit{e.g.}, Sec.~\ref{subsec:psa}, Fig.~\ref{fig:qpsa}).

The Si detectors and the  photodiode reading the CsI crystal where connected to PACI preamplifiers~\cite{paci}
placed under vacuum very close to the detectors.
Signals were then brought outside vacuum using 8~m long differential cables,
and connected to custom made digitizers. The digitizers for Si detectors, already used in
all previous FAZIA tests, feature 14 bit ADC's with 100 MHz sampling rate.
The effective number of bits (ENOB) of each digitizing board is about 11.4.
Signals from the photodiode were sampled by a 12 bit/125 MHz digitizer~\cite{pasquali_dsp}.

In this work information coming from the CsI detector will be  used only  
for vetoing particles not stopped in Si2 (punch-through particles). The CsI has a tapered
shape well suited for a distance of 1 m from the target: at such a distance, particles
 passing through the Si detectors will be completely contained in the CsI in spite
of their diverging paths. However, at the employed distance of 50 cm, particles
impinging near the borders of the Si can escape the CsI. Therefore the CsI veto is
not 100\% efficient, as it will be clear from the \dee and \psa correlations showed
in Figs.~\ref{fig:dee} and~\ref{fig:qpsa}.

The acquired signal length for Si1 is 20$\,\mu$s. For Si2,
since charge collection times in an underdepleted detector can be more than an order
of magnitude longer than at full depletion, the length of acquired signals has been
set to 70$\,\mu$s (\textit{i.e.} 7000 samples,  maximum length allowed by the
FEE signal memory).

\begin{table}[htbp]
 \begin{center}
\begin{tabular}{c c c c c}
\hline 
\\
Voltage   & Depletion  &   Undepl.  & Max. Rise  & Acquired\\
on Si2  &  Depth &   Layer &  Time (20-70\%)  & Signals \\
(V) & ($\mu$m) &   ($\mu$m)&   ($\mu$s) & ($\times 10^5$)\\
\\
\hline \\

105 & 310 & 200 & 13 & 9 \\
130 & 340 & 170 & 10 & 6 \\
200 & 420 & 90 & 3.0 & 5 \\
235 & 460 & 50 & 1.5 & 7 \\
290 & 510 &  0 & 0.45 & 7\\
\\
\hline
\end{tabular}
\end{center}
 \caption{}
 \label{tab:Si2bias}
\end{table}

\begin{figure}[htbp]
  \centering
   \includegraphics[width=9cm]{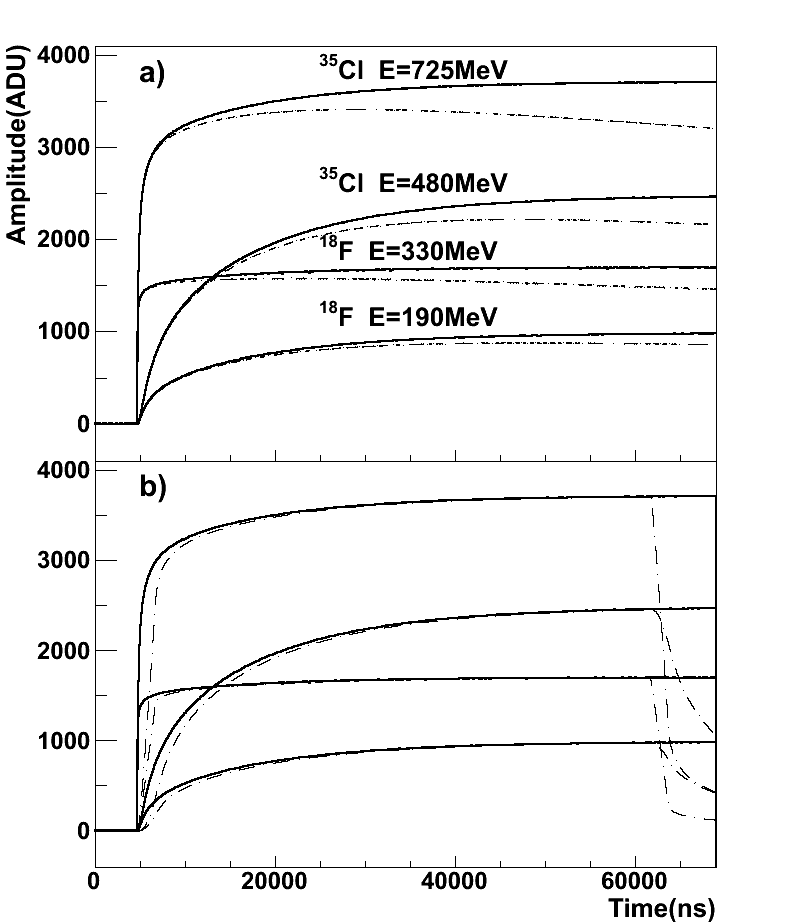}
   \caption{Panel a): Digitized preamplifier signals  in Si2 for  $^{18}$F at 190  and 330~MeV  and for
 $^{35}$Cl at 480  and 725~MeV. Each ion is stopped in the undepleted region at the lower energy
   and in the depleted region at the higher energy. The ranges associated with the two energies are
     195~$\mu$m and 480~$\mu$m for $^{18}$F, 190~$\mu$m and 360~$\mu$m for $^{35}$Cl.
 For each energy, signals both with (full lines) and without pole zero cancellation (dashed lines)
are plotted.  Panel b): Preamplifier signals, corrected with  pole zero cancellation before (full lines) and after (dashed lines)
    trapezoidal shaping with a very long ($>$ 50~$\mu$s) flat top.
  Data in  both panels
   refer to the lowest employed bias voltage of 105~V.}
   \label{fig:signals}
\end{figure}

\section{Data Analysis} \label{sec:anal}

\subsection{Amplitude Measurement} \label{subsec:ampli}


For charge collection times of the order of 10$\,\mu$s or more, the decay of the charge preamplifier
signal (decay time constant $\approx\,$425$\,\mu$s) and the associated ballistic 
deficit~\cite{baldinger, loo_ballistic}
of the preamplifier signal (even before shaping) cannot be neglected (see Fig.~\ref{fig:signals}).
Moreover, the ballistic deficit would depend on signal rise-times, varying greatly with the energy and
the atomic number of the detected fragment, thus spoiling the energy resolution.
To minimize this effect, a pole-zero cancellation algorithm has 
been applied as  part of the waveform shaping. Figure~\ref{fig:signals}, panel a), 
shows examples of the acquired preamplifier signal for Si2 at the
lowest bias employed in the experiment (105~V).  
Signals refer to $^{18}$F and $^{35}$Cl ions. For each ion type, signals at two different energies are shown,
after baseline subtraction~\cite{bardelli_baseline}:
the lower energy corresponding to an ion stopped in the undepleted region, the higher energy
to an ion reaching the depleted region. The slow rise-time associated with the former, when compared to the 
latter, is apparent. 
     Dashed and full lines show the shapes of the original signals and of the signals corrected
     for the finite preamplifier decay constant (pole-zero cancellation), respectively.
     For a given particle and energy, signals with and without correction are almost coincident during the leading edge, 
     though their maximum amplitudes are substantially different.
A ballistic deficit due to the finite preamplifier decay constant clearly appears as a reduction
of a few \% (about 10\% for $^{35}$Cl) in the maximum signal amplitude without correction.

After baseline subtraction and pole-zero cancellation
the charge signal is shaped 
using a trapezoidal shaper algorithm. The shaper acts as a pass-band filter, with a bandwidth depending on
the leading edge rise-time (fixed at 2$\,\mu$s). The so called ``flat-top'' length must be long enough to reduce
ballistic deficit effects (\textit{i.e.} the step response of the shaper must last long enough to accomodate the whole
charge collection time). 
In Fig.~\ref{fig:signals}, panel b), the same preamplifier signals of panel a) with pole zero cancellation applied (full lines)
are shown, together with the corresponding trapezoidally shaped signals (dashed lines).  
A quite long flat top of 55$\,\mu$s has been used for Si2.
For ions stopped in the undepleted region (slowest signals) it is clear 
that  a shorter flat top of the trapezoidal shaper would 
produce  a supplementary ballistic deficit. For those particles, the flat top of the shaper (which
  has unitary gain) would not last long enough to reach the
  maximum amplitude of the preamplifier, which occurs more than 50 $\mu$s
  after the leading edge.
On the other hand, a 1$\,\mu$s long flat top is well suited for Si1 signals (not shown in Fig.~\ref{fig:signals}).
The energy information is obtained from the maximum amplitude of the shaped signal. Numerical simulations of 
the behaviour of our trapezoidal shaper, employing realistic input signals, showed that the ballistic deficit
  with a 55$\,\mu$s flat top is about 0.5\% for the maximum measured
rise-time ($\approx$ 13~$\mu$s, see Table~\ref{tab:Si2bias} and Fig.~\ref{fig:qpsa}) and less than 0.1\% for rise-times $<$ 8~$\mu$s..

\begin{figure}[t!]
 \centering
\includegraphics[width=9cm]{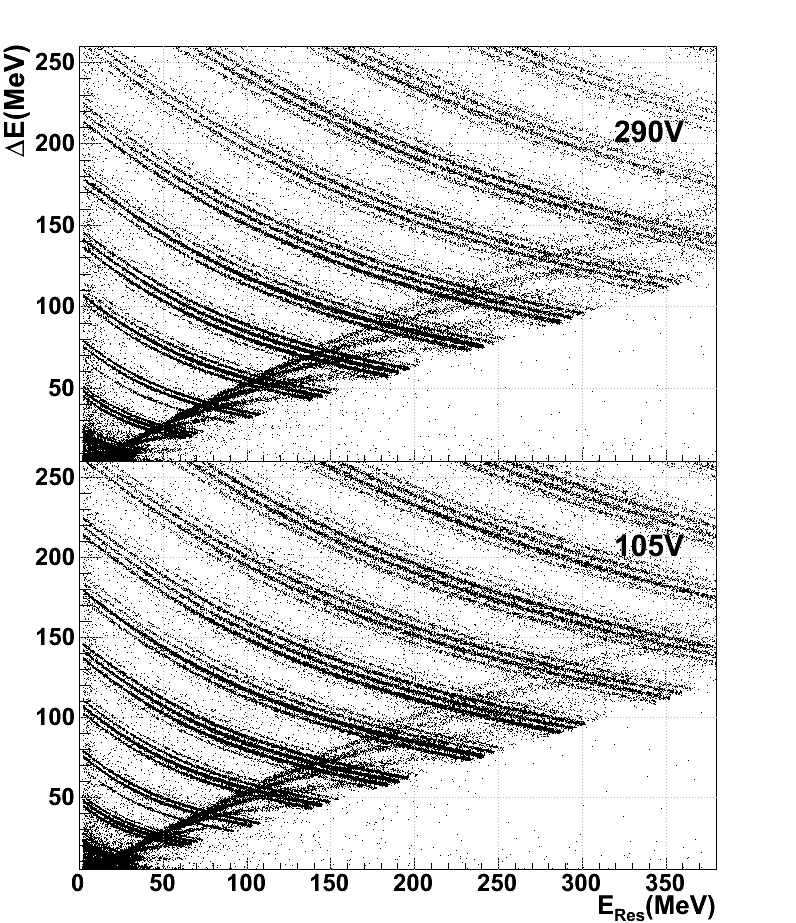}

 \caption{Top panel: \dee correlation ``Si1 vs Si2''  at a 290 V bias voltage.
Bottom panel: The same \dee correlation   at a 105 V bias voltage. Particles punching-through
Si2 are excluded using the CsI information: however, a residual contamination by such events is still
present due to geometrical reasons (see Sec.~\ref{sec:setup}).}
 \label{fig:dee}
\end{figure}

\subsection{\dee identification} \label{subsec:dee}


The \dee technique allows for isotopic identification of fragments stopped in Si2.
Already during data taking, a surprising agreement of the Si1-Si2 \dee correlations at all  applied
bias voltages became apparent, possibly suggesting that the charge collection efficiency is almost
independent (within few~\%) of the bias conditions (see Sec.~\ref{subsec:energy}).
Figure~\ref{fig:dee} shows, as an example, two \dee correlations (Si1 vs Si2) obtained at
the highest  (top panel) and lowest (bottom panel) bias voltages applied to Si2.
No degradation of the isotopic separation can be spotted in the figure, though at
105~V Si2 is only depleted by 60\%.

\begin{figure}[t!]
 \centering
\includegraphics[width=9cm]{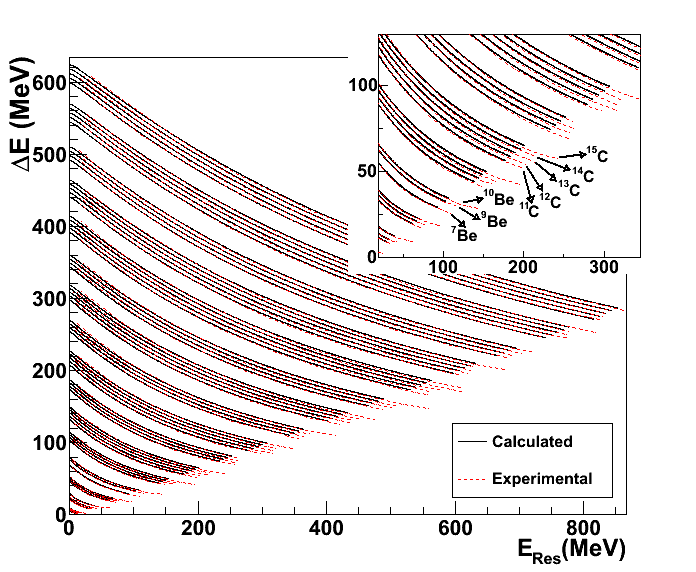}
 \caption{(color online) Comparison of the interpolation lines, manually drawn following the experimental isotopic ridges at 290~V,
(full lines)
with the curves obtained from energy loss calculations~\cite{hubert} (dashed lines). 
 The inset shows a detail of the plot in the region of light elements.}
 \label{fig:isotopic_id}
\end{figure}

A linearization procedure based on identification curves manually drawn on the ridges of the \dee correlation 
 allows for extraction of the so-called Particle IDentification (PID) parameter. Intervals of PID are then assigned to a
definite (Z, A) pair. 
The assigment of
the correct (Z, A) pair to a given \dee ridge is mainly a matter of self-consistency.
The Z value can be easily derived just by counting the ridges in the \dee correlation (the 
Z=4 lines are easily recognized since $^8$Be decays before reaching the detector, leaving
a gap between $^7$Be and $^9$Be). 
The mass assignment is an easy task for the lighest fragments with $Z\le4$.
For $Z>4$ a wrong mass value assigned to one of the isotopes employed in the calibrations (see  Sec.~\ref{subsec:energy})
would produce a sizeable increase of the $\chi^2$ of the calibration fit. 
Mass values are   assigned first to the isotopes employed for calibrations at 290~V. The stability of the \dee
correlation with applied voltage guarantees that the same assignements are also valid at the other
voltages. Extension to other isotopes, not employed for calibration, is  obtained by comparing 
the experimental \dee correlation,
after energy calibration, with the estimates of the energy loss calculations.
Figure~\ref{fig:isotopic_id} illustrates the good agreement between experimental data and
energy loss estimates: the full lines are the interpolation lines manually drawn on top of the
experimental ridges in the framework of the linearization procedure. The dashed lines
are obtained from energy loss calculations exploiting the range-energy tables of~\cite{hubert}.



%


%

\begin{table}[htbp]
 \begin{center}
\begin{tabular}{l | l l l l l}
\hline 
\\

    & \multicolumn{5}{ c }{Applied Voltage (V)}\\
Isotopes     & 105   &   130   & 200  & 235 & 290\\
\\
\hline \\

$^6$Li-$^7$Li       & 1.51(9)    & 1.5(2)  & 1.4(1)  & 1.5(1)  & 1.5(1)  \\
$^{11}$B-$^{12}$B   & 1.52(4)    & 1.50(6) & 1.54(6) & 1.43(6) & 1.42(5) \\
$^{12}$C-$^{13}$C   & 1.50(5)    & 1.44(5) & 1.50(6) & 1.50(5) & 1.51(6) \\
$^{21}$Ne-$^{22}$Ne & 1.18(9)    & 1.21(6) & 1.20(8) & 1.18(3) & 1.19(8) \\
$^{24}$Na-$^{25}$Na & 1.10(2)    & 1.12(3) & 1.11(3) & 1.14(3) & 1.16(8) \\
$^{25}$Mg-$^{26}$Mg & 1.03(7)    & 1.1(1)  & 1.12(8) & 1.08(7) & 1.1(1)  \\
$^{29}$Si-$^{30}$Si & 0.98(3)    & 0.98(3) & 0.92(3) & 0.99(3) & 0.97(3) \\
$^{33}$S-$^{34}$S   & 0.63(6)    & 0.74(9) & 0.98(7) & 0.69(9) & 0.78(8) \\
\\
\hline
\end{tabular}
\end{center}
 \caption{Figure of Merit (FoM) for adjacent peaks in a PID (Particle IDentification) spectrum. 
 Events are integrated over all particle energies.
The first column reports the selected isotopic pair, the other columns  the associated FoM obtained at the different
bias voltages applied to Si2.
} 
 \label{tab:dee_fom}
\end{table}

As in previous FAZIA experiments (see, \textit{e.g.}, Ref.~\cite{carboni_lns09}), 
 the  ``Figure of Merit'' (\fom)
has been employed in order to quantitatively express the obtained discrimination power. 
From the PID spectrum, the \fom~\cite{fom} for adjacent isotopes is defined as:
\begin{equation}
 \mathrm{FoM} = \frac{|\overline{\mathrm{PID}_1}\,-\,\overline{\mathrm{PID}_2}|}{\mathrm{FWHM}_1\, +\, \mathrm{FWHM}_2}
\end{equation}
where $\overline{\mathrm{PID}_1}$ and $\overline{\mathrm{PID}_2}$  are the centroids of the peaks associated to
two neighboring isotopes and $\mathrm{FWHM}_1$ and  $\mathrm{FWHM}_2$ are their full
widths at half maximum.
Table~\ref{tab:dee_fom} reports FoM values, integrated on energy, for fragments stopped in Si2.
For a given isotopic pair, similar \fom values are
found at all the employed bias voltages applied to Si2. This
proves that the \dee isotopic identification is not substantially affected by the partial depletion of
Si2, provided that the signals, which become very slow, are suitably treated, as
described in sect.~\ref{subsec:ampli}.


\subsection{Energy Calibration} \label{subsec:energy}

A usual calibration technique for  \dee telescopes exploits the so-called ``punch-through'' points of different isotopes
(see Ref.~\cite{braunn} $\oint$ 3.2 or Ref.~\cite{carboni_lns09} $\oint$ 2.2).
These are the  points with the largest residual energy \eres of each \dee curve.
Punch-through points in ADC units (ADU) have
been estimated by visually inspecting the \dee correlations in the punch-through region and  the error on 
these  values has  been estimated to be usually less than 1\% for
\de and 1-2\% for \eres.
From energy-range tables like~\cite{hubert} it is then possible to associate the proper deposited energies
to the uncalibrated values. For each silicon detector, 37 punch-through points (for different isotopes with $2\le$Z$\le16$)
have been used in order to calibrate the energy scale. A simple proportionality  between ADC units and
MeV has been assumed, as in previous FAZIA experiments. 
 Table~\ref{tab:calib} summarizes the calibration fit results for Si2 at
the different bias voltages. The calibration factor (second column of Table~\ref{tab:calib})
 is determined with an uncertainty of less than 0.1\% and it
decreases steadily, for Si2, going from 105 to 290V. However, it changes at most by 2\% from 
the lowest to the highest applied voltage. 

From the reduced $\chi^2$ values (fourth column of Table~\ref{tab:calib}) a reasonable agreement of the data with the assumed 
proportionality can be inferred. 
Since $\chi^2$
increases from 290~V to 105~V, a small deviation from the assumed proportional law
is suspected to be present at the lowest voltages.

%

\begin{table}[htbp]
 \begin{center}
\begin{tabular}{c c c c c c}
\hline 
\\
Voltage   & Conv.  &   $\chi^2$  &  $\chi^2_R=$    & $\delta_{290\mathrm{V}}$ & Si-eq.\\
on Si2    &  Factor     &             &   $\chi^2$/DoF  &                          & dead-layer    \\
(V) & ($\frac{\mathrm{MeV}}{\mathrm{a.u.}}$) &     &       & \% & ($\mu m$)    \\
\\
\hline \\

105 & 0.1962 & 34.5 & 0.96  &  2.03  & 17\\
130 & 0.1949 & 33.9 & 0.94  & 1.35 & 11\\
200 & 0.1937 & 26.3 & 0.73  & 0.73  &  6  \\
235 & 0.1930 & 16.7 & 0.46  &  0.36 & 3\\
290 & 0.1923 &  24.3 & 0.68  & \textemdash& 0\\
\\
\hline
\end{tabular}
\end{center}
\caption{First four columns: 
applied voltages, calibration factors, associated $\chi^2$ and reduced $\chi^2$ values for Si2. 
In all cases, the number of degrees of freedom (DoF) is 36.
The standard deviation of the reduced $\chi^2$ is 0.24. Last two columns: 
relative difference of calibration factors with respect to 290~V bias and estimated Si-equivalent
entrance dead-layer (see Sec.~\ref{subsec:energy}).} 
 \label{tab:calib}
\end{table}

In the undepleted region one expects a very low, if not zero, electric field. Therefore,
apart from the slowing down of charge collection already shown in Fig.~\ref{fig:signals},
one would also expect a reduction in collection efficiency with respect to full depletion. 
While the general slowing down of charge collection is substantially
confirmed (see also Sec.~\ref{subsec:psa}), the reduction in collection efficiency
(reported in the fifth column of Table~\ref{tab:calib})
 seems to be  at most   $\sim$2\% at the lowest bias voltage. This value
 points to a high collection efficiency from the undepleted region.
However, the calibration factors were obtained exploiting particle tracks extending for the whole detector thickness.
Particles with shorter range
will be addressed in Sec.~\ref{subsec:calib}.

We remind that  it is essential to use very long shaping times 
and to apply pole-zero cancellation of preamplifier decay  (see Sec.~\ref{subsec:ampli})
to achieve such a result. 
Otherwise one observes the well known
ballistic deficit, an effect related to electronics and independent of charge collection efficiency~\cite{baldinger, loo_ballistic}.
On the other hand, ballistic deficit cannot explain the observed 2\% variation of the calibration factor,
since it has been estimated to be less than 0.5\% (see Sec.~\ref{subsec:ampli}).

One can convert the observed dependence of the calibration factors on bias voltage into 
 a Si-equivalent effective dead layer at the entrance side. Assuming a zero dead layer at 290~V,
we calculated the dead layer thicknesses which would give the observed calibration factor
at the other bias voltages. They are reported in
the rightmost column of Table~\ref{tab:calib}. However, a real dead layer at the entrance of Si2
would affect not only the measured energies
at the punch-through in Si2: it
would affect even more the energy measured by  Si1 at the leftmost point
of the \dee curve, when the particle begins to give a signal in Si2.
That energy in Si1, which is easily derived by inspection of the \dee correlations,
is practically the same for all the employed biases
and it is not compatible with the dead layer thickness quoted in  Table~\ref{tab:calib}.
In fact, a reduction of several MeV should be observed  in presence of a dead layer of the estimated thickness.
For example, a dead layer of 17~$\mu$m Si-equivalent estimated at 105~V would result in a reduction of
about 10~MeV for the energy deposited in Si1 by a $^{10}$B, while
that value stays constant within 1~MeV.
It is therefore possible to exclude the presence of a totally inactive layer at the entrance  of Si2.

\subsection{Energy response of ions as a function of the bias voltage}  \label{subsec:calib}

The results of Sec.~\ref{subsec:energy}, showed that for particle tracks extending over the whole
detector thickness
the ``end points'' of each \dee curve are stable in amplitude, within 2\%, as a function 
of the bias voltage. A deeper investigation was devoted to studying the detector response in the whole
energy range spanned by the \dee correlations. 
To avoid  systematic effects due to the energy loss calculations, we compare the charge-signal amplitude
at different bias voltages with that at full depletion (taken as a reference), being
the signal amplitude  directly related to the collection efficiency.
The detector response for each specific fragment type can be derived, thanks to the
isotopic identification 
allowed by the $\Delta$E(Si1) vs $\mathrm{E_{RES}}$(Si2) correlations
 (see Sec.~\ref{subsec:dee}).

%
%

For a given identified isotope,  the events have been classified in bins of $\Delta$E(Si1).
Bins having the same  $\Delta$E(Si1), though taken at different bias voltages applied to Si2, correspond
to the same incident energy (and to the same range in Si2, for a given A and Z). As a consequence, 
they also correspond to the same value of \eres physically deposited
in Si2: charge amplitudes recorded for such events are thus directly comparable.
 
\begin{figure}
 \centering
\includegraphics[width=9cm]{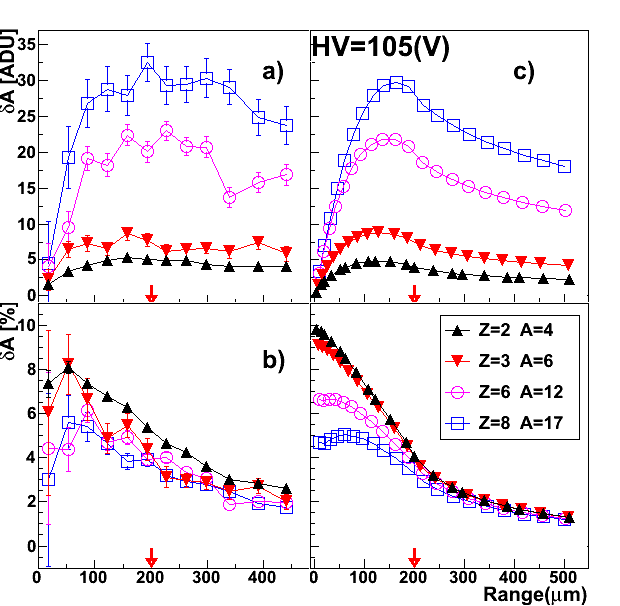}

 \caption{
(color online) Panel a): experimental amplitude difference (ADC units) between events at 290~V and events at 
105~V bias voltage as a function of particle
range (estimated using the \de information from Si1). Data refer to four different isotopes: $^4$He, $^6$Li, $^{12}$C and
 $^{17}$O; panel b): same data as in panel a) except that the amplitude difference is given as a percentage of the amplitude
at 290~V;
  panel c) calculated difference in the total collected charge, 
 to be compared with the data in panel a) (see Sec.~\ref{subsec:calib} for details); panel d)
 same values as in panel c),  reported as a percentage of the total collected charge for comparison with panel b).
The undepleted region is  200 $\mu$m thick: this value is evidenced by the arrow in each panel. The error bars
refer to the statistical uncertainties.}
 \label{fig:amp_diff105}
\end{figure}

Panel a) of Fig.~\ref{fig:amp_diff105} shows, as a function of estimated
particle range, the amplitude difference in ADC units (ADU)
$$\delta\mathcal{A}=\mathcal{A}_{\mathrm{Si2}}(290\,\mathrm{V})-\mathcal{A}_{\mathrm{Si2}}(105\,\mathrm{V})$$
The same difference is also reported in panel b) as a percentage of the value of $\mathcal{A}_{\mathrm{Si2}}(290\,\mathrm{V})$.
Data refer to a few nuclear species, namely $^4$He, $^6$Li, $^{12}$C and $^{17}$O.
Similar results are obtained at a bias voltage of 130~V (not shown); only the
absolute and relative differences are reduced by about a factor of 2.
At a bias of 200~V or greater, the difference $\delta\mathcal{A}$ has been found compatible with
zero within the  errors.


The range, \textit{i.e.} the penetration depth in the detector, revealed itself as the
most relevant parameter to study the detector response. As a matter of fact,  the amplitude
difference at the two lowest biases increases up to a range approximately equal to the undepleted zone thickness 
(evidenced by the  arrow) for all fragments.
After that point, the difference remains about constant or slightly decreases.
Relative differences of at most 8\% (cf. $^4$He at 105~V, Fig.~\ref{fig:amp_diff105} panel b)) can be noticed.

As  expected, for ranges of the order of the detector thickness the relative difference (Fig.~\ref{fig:amp_diff105}, panel b))
reaches the
 values obtained from the calibration factors in Table~\ref{tab:calib}
for the same bias voltages (2\% for 105~V).

 A detailed theoretical study of the collection process would be well
beyond the aim of the present paper. Realistic numerical simulations, as those presented in~\cite{bardelli_sim, sosin, parlog2010, hamrita2011},
would be an ideal tool for such a study.
Here we  simply attempt an  empirical description, trying to explain the experimental data by assuming
 incomplete charge collection from the undepleted region.
The measured amplitude is proportional to the collected charge. The
charge carriers per unit thickness along the track are proportional, as a function of the pe\-ne\-tra\-tion depth $x$,
 to the Bragg curve for the given fragment.
Therefore, the contribution to the final amplitude coming from a given segment of the track
is proportional to the integral of the Bragg curve over the same interval.
To take into account the incomplete charge collection, each segment of the Bragg curve should be
weigthed by an ``efficiency factor''. 
We ideally divide
 the detector into a depleted and an undepleted region and we
apply different efficiency factors to the two regions, assuming 100\% efficiency in the depleted region.
It is apparent that a constant collection efficiency across the undepleted region cannot
explain data of Fig.~\ref{fig:amp_diff105}  panel b). In fact, for particles stopped in the undepleted region,
it would  give a constant relative difference independent of particle range.
Assuming a variable collection efficiency $\eta(x)$, the simplest hypothesis is that of a linear
variation with the distance $x$ from the entrance surface, starting at some $\eta(0)<1$ value and
reaching $\eta(d)=1$ where $d$ is the thickness of the undepleted region ($\eta=1$ all over the depleted region).
 Moreover,
Fig.~\ref{fig:amp_diff105} shows that the charge collection is
more efficient for heavier fragments,
\textit{i.e.} those having a higher ionization density for a given penetration depth. Therefore, 
 we introduce a term dependent on the stopping power $\left|\mathrm{d}E/\mathrm{d}x\right|$
in the collection efficiency
(assuming a linear dependence for the sake of simplicity), obtaining 

\begin{equation*}
\eta(x)=\begin{cases}
        \eta(0)+(1-\eta(0))\frac{x}{d}\,+\,\alpha\left|\frac{\mathrm{d}E}{\mathrm{d}x}\right|\frac{d-x}{d}, & \text{if $x < d$.}\\
       1, & \text{if $x \ge d$.}       
\end{cases}
 \end{equation*}

For each fragment and each value of the range, the  ``measurable'' deposited energy
is obtained as:

$$E=\int \left|\frac{\mathrm{d}E}{\mathrm{d}x}(x)\right|\,\eta(x)\, \mathrm{d}x$$

Panels c) and d) in Fig.~\ref{fig:amp_diff105}
present the result of our calculation. Since the integral of the Bragg curve gives an energy,
we have converted it to ADU, exploiting the calibration factors of Table~\ref{tab:calib}. 
A reasonable agreement with the experimental
values has been obtained for $\eta(0)=0.89$ ($\eta(0)=0.92$)  at 105~V (130~V) and $\alpha=0.4$~$\mu$m/MeV.

The presented phenomenological approach seems to contain the right ingredients to reproduce the
experimental behaviour. It has not been obtained  from first principles or
a microscopic description of the charge collection process and  it is thus unable to give us
detailed physical information. However, it can be used to get an approximate value of the
charge collection efficiency in the undepleted region. 
On the whole, one can say that a maximum collection deficit of about  10\% in the undepleted 
region is compatible with the observed behaviour for all reported fragments at the two lowest bias voltages
of 105 and 130~V while data at 200 and 235~V are compatible with a collection efficiency of
about 100\%.
 To conclude, a surprisingly high average collection
efficiency from the undepleted region (90\% or more) must be assumed
to reasonably reproduce the data
at the two lowest biases. 

\begin{figure*}[t!]
 \centering
 \includegraphics[width=16cm]{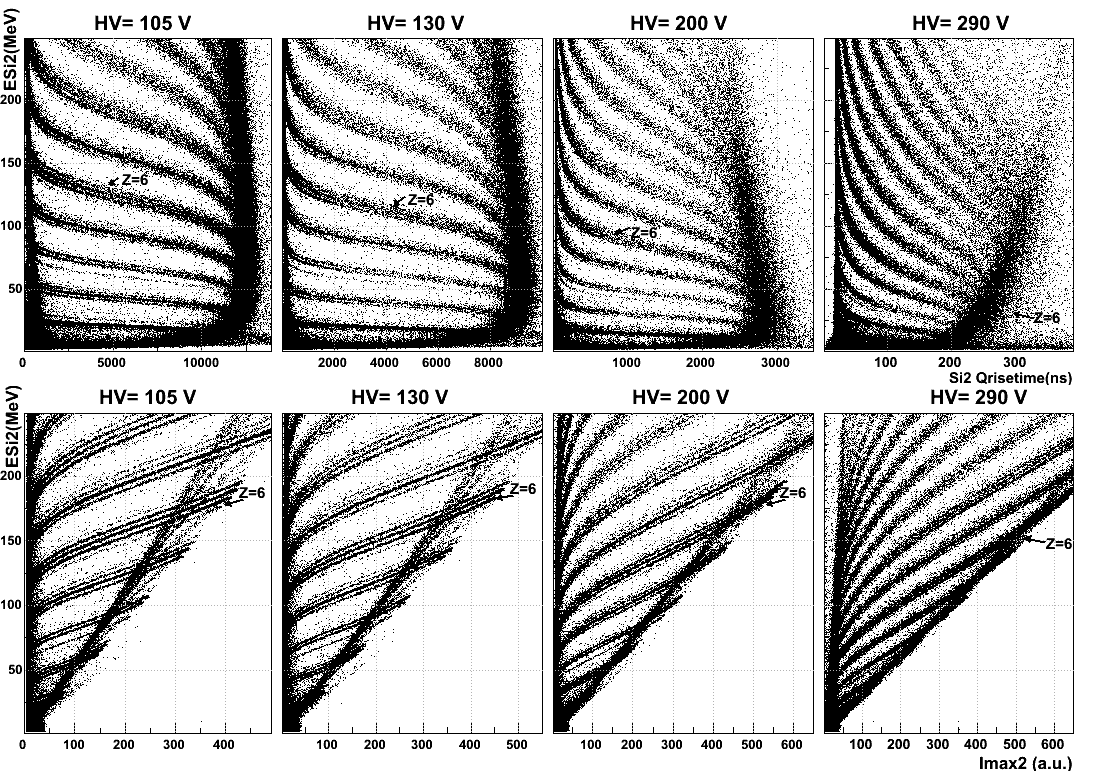}
 \caption{Top row: ``Energy vs Charge rise-time'' correlations  at different bias voltages.
Bottom row: ``Energy vs Current Maximum'' correlations  at the same bias voltages.}
 \label{fig:qpsa}
\end{figure*}

From the observed behaviour, assuming a linear energy response at full depletion, 
a non-linearity of the order of few \% can be inferred. In principle, this effect
could be corrected for by exploiting the very same data shown in Fig.~\ref{fig:amp_diff105}.
 A preliminary analysis shows that
a simple second order polynomial correction permits to
obtain the ``full depletion'' amplitude (\textit{i.e.} the value which would be obtained at 290~V)
from the experimental amplitude at lower bias voltage,
once the fragment has been correctly identified using either the PSA or the \dee technique.
However, the coefficients of the polynomial are different for different isotopes and it is still not clear
if they can be derived from a simple functional dependence on Z and A.

\subsection{Pulse Shape Analysis}  \label{subsec:psa}

The PSA technique permit to identify particles stopped in one
silicon detector from information delivered by that detector
alone. PSA will thus allow to reduce the identification thresholds when applied
to the first telescope stage in a physics experiment. 
All results presented in this section have been
obtained with a veto condition on the CsI(Tl) detector to select particles
stopped in the detector under test, which in this work is Si2.
Two methods of PSA have been used, based on the two correlations ``Energy vs Charge
rise-time'' and ``Energy vs Current maximum''. Both techniques
had been already investigated within the FAZIA R\&D program~\cite{bardelli_lnl07,
carboni_lns09, leneindre_lns11}.

\begin{figure*}
 \centering
 \includegraphics[width=14cm]{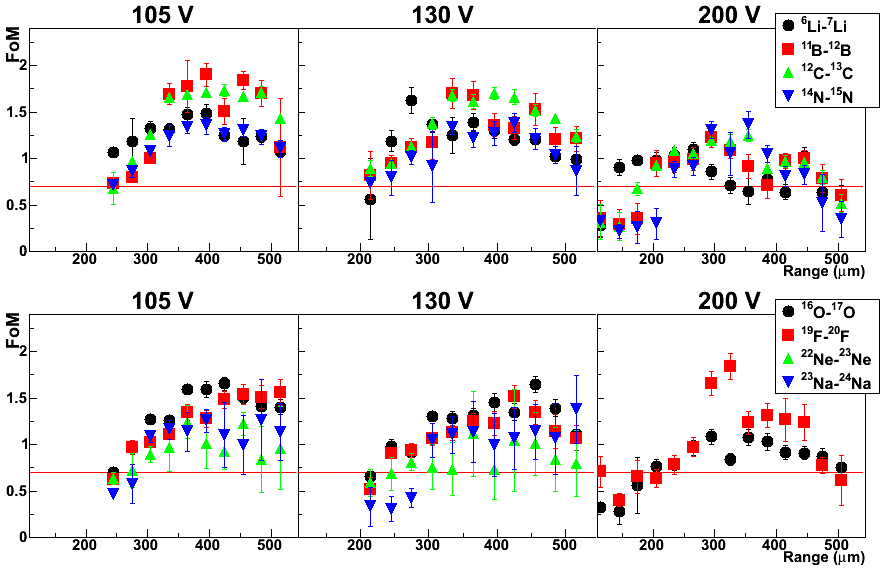}
 \caption{(color online) Isotopic identification with PSA using the ``Energy vs Current maximum'' correlation. 
   The Figure of Merit is shown for several isotopic pairs at 105, 130 and 200~V bias voltage as a function of particle range.}
 \label{fig:fomI}
\end{figure*}

In the top row of Fig.~\ref{fig:qpsa} the correlation ``Energy vs Charge
rise-time'' is shown for Si2 at four of the five bias voltages employed in this work,
namely 105, 130, 200 and 290~V: isotopic separation is quite good
at 105~V and  it worsens with increasing bias voltage,
eventually disappearing for an applied voltage greater than 200~V (the correlation at
235~V, not shown in the picture, is similar to that at 290~V). The span of the horizontal axis
shows how the rise-time decreases from about 13~$\mu$s to less than 400~ns when
going from 105 to 290~V bias voltage.

Already from the top row of Fig.~\ref{fig:qpsa} one can get an estimate of the minimum energy for which isotopic
separation is possible. For instance, a threshold of about 110~MeV can be inferred for carbon (Z=6) both
at 105 and 130~V. 

The second PSA method is the so-called ``Energy vs Current maximum'' technique. 
The bottom row of Fig.~\ref{fig:qpsa} shows
the correlation ``Energy vs Current maximum'' for Si2 at the same voltages as
the top row. A better performance of this technique with respect to
the previous one  can be inferred.
However, as for the ``Energy vs Charge
rise-time'' correlation, the isotopic separation worsens with increasing bias voltage. 
A residual contamination of punching-through particles, not vetoed by the CsI for geometrical reasons
(see Sec.~\ref{sec:setup}), is  present. They form ridges approximately along the
diagonal in  the ``Energy vs Current maximum'' correlation (from the bottom-left corner to the
top-right one).

The isotopic discrimination limit can be quantitatively evaluated by means of the already defined
\fom~\cite{fom}, after the ``Energy vs Charge rise-time'' 
or the ``Energy vs Current Maximum'' correlations have been linearized by extracting a PID value,
as already explained for \dee in sect.~\ref{subsec:dee}. 
In this work we focus on the ``Energy vs Current Maximum''  technique since it seems to perform 
better than the ``Energy vs Charge rise-time'' one. In fact, 
\fom values extracted from ``Energy vs Current maximum'' correlations, are higher than
those obtained from the   ``Energy vs Charge rise-time'' method.


Figure~\ref{fig:fomI} shows the  \fom  obtained from ``Energy vs Current maximum'' correlations
as a function of the particle range. 
%
We  consider two isotopes as ‘‘well separated’’ if \fom
is greater than 0.7 (see ref.~\cite{bardelli_lnl07}), corresponding to a peak-to-valley ratio of 2 when 
the two peaks are equal.
For each isotopic pair a minimum penetration depth for which \fom$>$ 0.7 can be inferred from graphs
like those shown in Fig.~\ref{fig:fomI}: it constitutes the lower threshold, in range, for isotopic
identification. Such thresholds are shown in Table~\ref{tab:id_thr} for various isotopic pairs
at the different bias voltages. The corresponding incident energy is also reported. 
Figure~\ref{fig:fomI} also shows that in a few cases
the \fom value decreases for higher penetration, thus falling below the \fom$=$0.7 line: for those cases
we give range (and energy) intervals for good separation instead  of a lower threshold. 
The expectation that the penetration depth is a more useful parameter than energy is confirmed by 
the fact that the identification thresholds (crossing of \fom=0.7) in range are about the same for all Z values
at a given bias voltage. 
The energy thresholds are found in good agreement 
with those visually estimated from
the ``Energy vs Current Maximum'' correlations.
We notice that at 105~V and 130~V bias voltage the ``Energy vs Current maximum'' method permits
isotopic separation up to Al and Mg isotopes, respectively. In comparison, 
the ``Energy vs Charge rise-time'' method only allows isotopic 
separation up to F (at 105~V) or C (at 130~V).

\begin{table*}[htbp]
 \begin{center}
\begin{tabular}{l | l l |  l l  | l l}
\hline 
   &  \multicolumn{2}{ c }{105~V} & \multicolumn{2}{ c }{130~V} & \multicolumn{2}{ c }{200~V}\\
\hline
     Isotopes   &   Range ($\mu$m)   & Energy (MeV) &  Range ($\mu$m)   & Energy (MeV)  &  Range ($\mu$m)   & Energy (MeV) \\

\hline 
$^6$Li-$^7$Li       & 240-     &  40-      & 220-      &  40-     & 130-       &  30-    \\
$^{10}$B-$^{11}$B   & 240-     &  90-      & 220-      &  90-     & 170-       &  75-    \\
$^{12}$C-$^{13}$C   & 250-     & 120-      & 220-      & 110-     & 180-       & 100-    \\
$^{14}$N-$^{15}$N   & 250-     & 150-      & 220-      & 140-     & 200-480    & 130-225  \\
$^{16}$O-$^{17}$O   & 250-     & 190-      & 220-      & 170-     & 220-470    & 170-270  \\
$^{19}$F-$^{20}$F   & 250-     & 230-      & 230-      & 215-     & 230-470    & 215-335   \\
$^{22}$Ne-$^{23}$Ne & 260-     & 280-      & 240-      & 260-     & 240-390    & 260-355   \\
$^{23}$Na-$^{24}$Na & 270-     & 320-      & 290-400   & 335-410  &            &          \\
$^{26}$Mg-$^{27}$Mg & 280-     & 380-      & 310-380   & 400-460  &            &            \\
$^{27}$Al-$^{28}$Al & 320-410  & 455-530   &           &          &            & \\

\hline
\end{tabular}
\end{center}
 \caption{Range and energy intervals for which good isotopic identification (\textit{i.e.} \fom $>$ 0.7) is achieved.
Data refer to selected isotopes with $3\le\mathrm{Z}\le13$ and to three applied voltages: 105, 130 and 200~V.
A single value is reported when isotopic separation is achieved for stopped fragments of all ranges/energies above the low
threshold. No value is reported when
\fom $<0.7$ for all range/energies at the given bias voltage.   
} 
 \label{tab:id_thr}
\end{table*}

\begin{figure*}[t!]
 \centering
 \includegraphics[width=16cm]{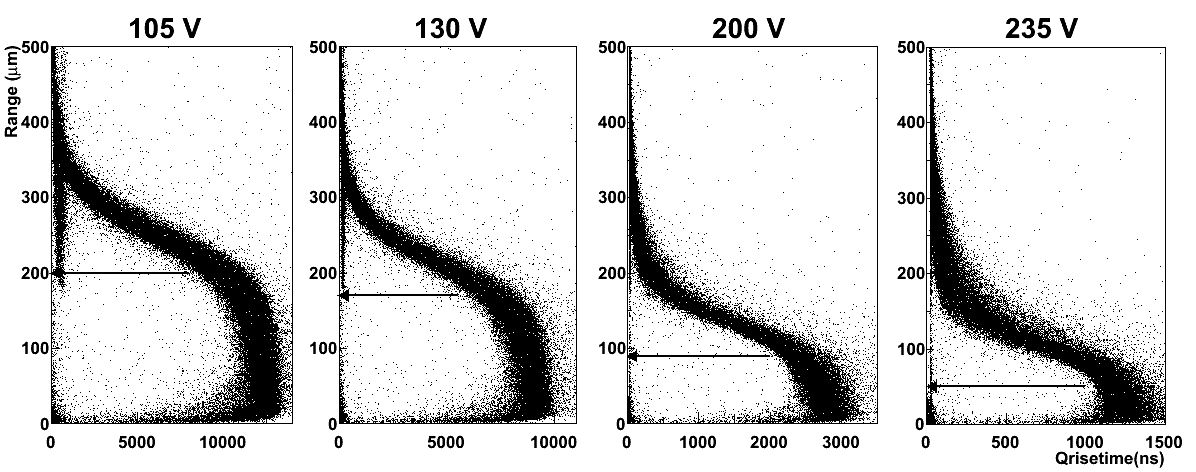}
 \caption{``Range vs Charge rise-time'' correlations at 105, 130, 200 and 235~V for fragments with Z$>$2. The arrows point to the
 estimated thickness of the undepleted region.}
 \label{fig:funnel}
\end{figure*}

One also notices that no isotopic identification is obtained for fragments stopped in the
undepleted region, \textit{i.e.} the \fom crosses 0.7 for range values greater than the undepleted layer
thickness. As a matter of fact, a rapid transition from a slow to a fast regime
of charge collection can be recognized for all fragments as soon as they get close to
the depleted region. This is particularly evident from the ``Range vs Charge rise-time''
correlations of Fig.~\ref{fig:funnel}, which refer to all fragments with Z$>$2. It is interesting to
see that the different ridges of Fig.~\ref{fig:qpsa}, top row, tend to collapse on a single ridge
when the fragment range is reported on the y-axis instead of the energy. Moreover, the rise-time stays
approximately constant for range values within the undepleted region and it starts decreasing at a
higher rate as soon as particles reach the depleted region, \textit{i.e.} the region of non-zero electric field.
This could  be a signature of
\textit{field-enhanced funneling}~\cite{funn1, funn2, funn3, funn4}, 
an extension of the electric field into the undepleted
region along the ion track, which can produce a faster charge collection.
A deeper study of this effect is planned by the FAZIA collaboration.

\begin{figure}
 \centering
 \includegraphics[width=9cm]{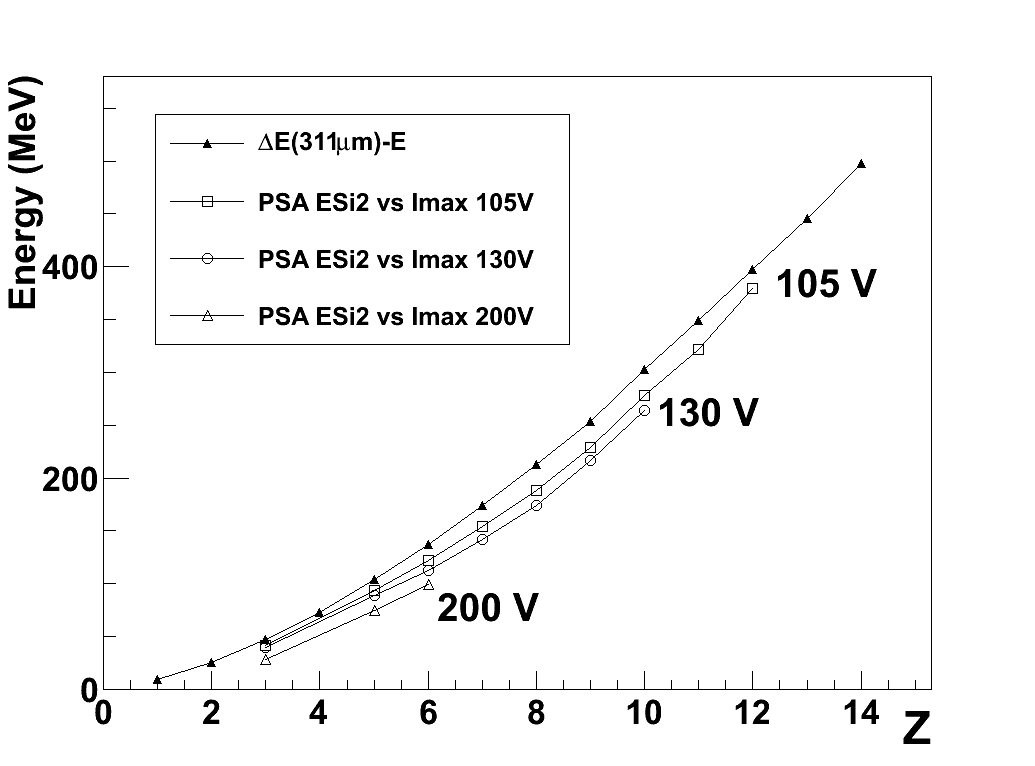}
 \caption{Energy thresholds for isotopic identification at various bias voltages: 105~V (empty squares), 130~V (empty circles) and
200~V (empty triangles).
The energy threshold for isotopic identification as given by the \dee technique for a 311 $\mu$m thick \de
detector
is also plotted (full triangles).}
 \label{fig:ethr_mev}
\end{figure}

In the spirit of Fig.~17 of Ref.~\cite{carboni_lns09}, the isotopic identification threshold
 in terms of deposited energy is plotted as a function of Z in Fig.~\ref{fig:ethr_mev}.
Different symbols in Fig.~\ref{fig:ethr_mev}
correspond to different bias voltages, namely 105~V (empty squares), 130~V (empty circles) and 200~V (empty triangles).
 In the same picture, for the sake of comparison, the energy threshold for isotopic identification as 
given by the \dee technique for a 311 $\mu$m thick \de detector
(\textit{i.e.} the minimum energy needed for the particle to punch through Si1 and to deposit some energy in Si2)
is plotted as full triangles.
As already shown in Table~\ref{tab:id_thr}, the threshold for mass identification increases for decreasing bias voltage. 







\begin{figure}
 \centering
 \includegraphics[width=9cm]{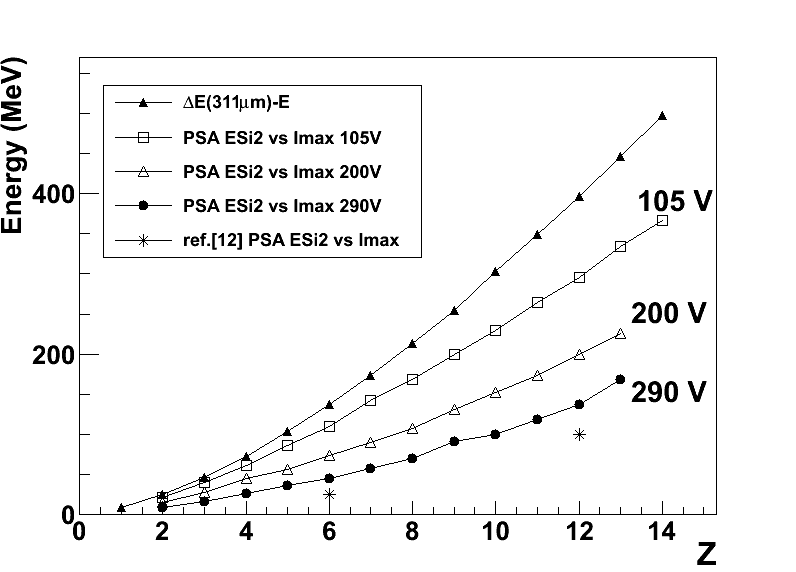}
 \caption{
Charge identification thresholds estimated from visual inspection of the
``Energy vs Current maximum''  correlations for a bias voltage of 105 (empty squares), 200 (empty triangles)
or 290~V (full circles). Thresholds affecting the \dee technique are also shown as full
triangles for a 311 $\mu$m thick detector. Values  from a previous \fazia work (Ref.~\cite{carboni_lns09}) 
are also plotted (stars, see text).}
 \label{fig:Zthr}
\end{figure}

For what concerns the charge identification,  
we will not deal with \fom values because a visual inspection of the 
``Energy vs Current maximum'' correlations is sufficient to determine the energy thresholds. 
From Fig.~\ref{fig:qpsa}, bottom row, 
it is apparent that at 105~V and 130~V bias voltage the energy threshold for charge identification
is slightly lower than that for mass identification. 
This energy threshold for charge identification, shown in Fig.~\ref{fig:Zthr} 
 at bias voltages of 105, 200 and 290~V for Z=2-13, becomes higher when reducing the bias voltage. This indicates 
that the better isotopic identification at low bias 
is achieved at the price of higher charge identification thresholds.
In order to compare the present results with those obtained by the collaboration in previous experiments, 
we also show the identification thresholds as reported in~\cite{carboni_lns09} 
for the `Energy vs Current maximum'' method (stars).
Since those data were  obtained with a fully depleted detector, it is no surprise that
the present result  at full depletion gets quite close to the previous data.
However, the  thresholds in~\cite{carboni_lns09} were slightly lower and
this could be due to the different  doping uniformity (better than 1\% in~\cite{carboni_lns09}, only  6\% for the present detector)\footnote{The 
different thickness of the detectors (306~$\mu$m in~\cite{carboni_lns09}, 510~$\mu$m in the present
work) should  make no significant difference, according to the previous FAZIA experience.}.

\section{Conclusions}


A study of the response of a 500 $\mu$m thick n-TD Si detector, mounted as the second stage of a Si-Si-CsI telescope
and  biased at a voltage below that necessary to obtain full depletion, has
been presented. The study takes into account both the energy response and the PS response of the detector.

For particles with a range equal to the detector thickness, the charge collection changes just by 2\%
between 105~V (60\% depletion) and 290~V (full depletion) bias voltage.
A remarkably high charge collection efficiency (about 90\% or more) from the undepleted region of the
detector is obtained, provided that pole-zero cancellation is applied and 
that signals are treated with a suitably long shaping time of a few tens of $\mu$s.
The charge amplitude response of the detector
has been found linear within few \% even at 105~V, when the undepleted region is  200~$\mu$m thick. 
Non linearities of the order of 5-8\% in the amplitude-energy response
have been
noticed for particles stopped in the undepleted region at 105~V bias voltage. These non-linearities
can be corrected for, knowing the particle atomic and mass number, by using a simple second
order polynomial correction whose coefficients, however, are different for different ion types.

The \dee performance is not affected by the incomplete depletion even when
40\% of the wafer is not depleted. 



The detector under test did not allow isotopic identification via PSA when biased at full depletion voltage.
In fact its doping uniformity is only about 6\%, while previous tests performed by the collaboration
showed that a doping uniformity of  about 1\% FWHM or better is needed for isotopic identification~\cite{bardelli_lnl07}.
It is not easy for manufacturers to provide ingots of $\approx$ 1\% doping uniformity.
The present result  shows that it is still possible to get  isotopic identification with PS techniques  from
 detectors of worse doping uniformity. Underbiasing the first stage of a \dee telescope, one can
get isotopic identification at lower energy than with the \dee technique alone.
However, with respect to a good uniformity Si detector, the better isotopic resolution comes at the
price of somewhat higher charge identification thresholds. Therefore, for experiments requiring an 
isotopic identification of fragments, either a comprimise must be found
between the two conflicting requirements or the beam time should be partly
devoted to measurements at low detector bias.

The long shaping times necessary for partially depleted detectors could impose
the acquisition of relatively long signals (7000 samples in the present case).
One could avoid memory or time limitations by employing decimation~\cite{dtsp} after sampling, thus reducing
the number of samples which must be acquired and processed.
Pile-up issues could still limit the use of this technique at high counting rates,
but the low beam currents at Radioactive Beam facilities should pose no such problems. 


\section*{Acknowledgements}

The authors would like to thank the LNS Superconducting
Cyclotron staff, in particular D.Rifuggiato, for providing a very
high quality beam. The support of the Machine shop teams of LNS,
of the Physics Departments of Florence (in particular M.Falorsi), of the Physics Departments of 
Naples and of the Physics Departments of 
Bologna (in particular M.Guerzoni and S.Serra) is gratefully acknowledged.
The effective collaboration of the Physics Department of
Florence (and in particular of E.Scarlini), 
of IPN (Orsay) and of LPC (Caen) is gratefully acknowledged.
We also thank A.Boiano, A.Meoli and G.Tortone (INFN-Naples) for their
invaluable assistance.
Thanks are also due to N.Zorzi of FBK(Trento) for cooperative support in detector development.
The research leading to these results has received funding from the European 
Union Seventh Framework Program FP7/2007-2013 under Grant Agreement n$^\mathrm{o}$ 262010 - ENSAR
and  from grants of Italian Ministry of Education, University and Research under 
contract PRIN 2010-2011.

\end{document}